\documentclass[preprint,prd,noshowpacs,nofootinbib]{revtex4}
\usepackage{amssymb}
\usepackage{amsmath}

\begin{document}

\title{Stokes' theorem, gauge symmetry and the time-dependent Aharonov-Bohm effect}

\author{James Macdougall}
\email{jbm34@mail.fresnostate.edu}
\affiliation{Department of Physics, California State University Fresno, Fresno, CA 93740-8031, USA}

\author{Douglas Singleton}
\email{dougs@csufresno.edu}
\affiliation{Department of Physics, California State University Fresno, Fresno, CA 93740-8031, USA}

\date{\today}

\begin{abstract}
Stokes' theorem is investigated in the context of the time-dependent Aharonov-Bohm effect -- the two-slit
quantum interference experiment with a {\it time varying} solenoid between the slits. The time varying
solenoid produces an electric field which leads to an additional phase shift which is found to exactly 
cancel the time-dependent part of the usual magnetic Aharonov-Bohm phase shift. This electric field arises
from a combination of a non-single valued scalar potential and/or a 3-vector potential. The
gauge transformation which leads to the scalar and 3-vector potentials for the electric field is
non-single valued. This feature is connected with the non-simply connected topology of the
Aharonov-Bohm set-up. The non-single valued nature of the gauge transformation function has interesting 
consequences for the 4-dimensional Stokes' theorem for the time-dependent Aharonov-Bohm effect. 
An experimental test of these conclusions is proposed. 
\end{abstract}

\maketitle
\section{Introduction}
In this work we investigate the interplay between Stokes' theorem and gauge symmetry
in the context of the {\it time varying} Aharonov-Bohm (AB) effect \cite{AB, ES}. Already the static 
AB effect -- placing an infinite magnetic flux carrying solenoid between the slits of a
quantum mechanical two-slit experiment -- shows a deep interrelation between gauge symmetry and
the 3-dimensional Stokes' theorem. If one allows the current and therefore the magnetic flux through the 
solenoid to be time dependent then one needs to take into account both the electric field (generated
from ${\bf E} = - \partial _t {\bf A}$) as well as magnetic field (generated from 
${\bf B} = \nabla \times {\bf A}$). The introduction of time dependence means that one 
must consider space-time coordinates and differentials ({\it i.e.} $x^{\mu} =(t, {\bf x})$
and $dx^{\mu} =(dt, d{\bf x})$) in doing the relevant integrals rather than simply spatial coordinates
and differentials ({\it i.e.} $x^i ={\bf x}$ and $dx^i = d{\bf x}$). One needs to consider Stokes' theorem in 4-dimensional 
Minkowski space-time. We will show that the electric field coming from the time-dependent 3-vector potential,
associated with the time-dependent magnetic flux in the solenoid, can be written entirely in 
terms of the 3-vector potential, ${\bf A}$, or in terms of a scalar potential, $\phi$, or some combination of the 
two. The relationship between these different ways of writing the electric field is through a gauge
transformation $A^{\mu} \rightarrow A^{\mu} + \partial ^\mu \chi$. However, the gauge transformation function,
$\chi$, and the scalar potential, $\phi$, are found to be non-single valued. This has interesting 
consequences for Stokes' theorem for this case. This analysis also shows that the phase shift coming
from the electric field exactly cancels the time varying part of magnetic AB phase {\it during the period when the 
potential is being varied with respect to time}. Thus, one can experimentally test the conclusions
in this work. Although there has been ample experimental confirmation of the static Aharonov- Bohm effect
\cite{chambers, tonomura} there has been no definitive experimental test of the time-dependent Aharonov-Bohm
effect to date. The one test that has been performed \cite{chentsov, ageev} supports the conclusions presented 
here that the AB phase shift does not inherit the time dependence of the magnetic flux.  
\section{Stokes' theorem using differential forms}
We begin by reviewing Stokes' theorem in 3 and 4-dimensions.
In 3-vector notation the 3-dimensional Stokes' theorem is given by 
\begin{equation}
\label{stokes3d}
\oint _{\partial S} {\bf A} \cdot d {\bf l} = \int _S \nabla \times {\bf A} \cdot d{\bf S} ~,
\end{equation}
where ${\bf A}$ is a 3-vector field and the first integral is a closed line integral 
and the second is an area integral over $S$ whose boundary is  $\partial S$. 
In differential forms notation Stokes' theorem takes the following elegant form
\begin{equation}
\label{stokes-df}
\oint _{\partial c} \omega = \int _c d \omega ~,
\end{equation}
(for a brief review of differential forms at the level needed here see reference \cite{ryder}).
In \eqref{stokes-df} $\omega$ is a $p$-form, $d \omega$ (the exterior derivative of $\omega$) is a
$p+1$-form, $c$ is a $p+1$ {\it chain} and $\partial c$ is the boundary of $c$, {\it i.e.} a
$p$ chain. In this work we will apply Stokes' theorem to electromagnetism so for our $p$-form $\omega$
we take the 1-form vector potential
\begin{equation}
\label{a-form}
\omega = A = A_\mu dx^\mu = \phi dt - {\bf A} \cdot d{\bf x} ~.
\end{equation} 
We have given the 1-form $A$ in 4-vector and 3-vector notation. Throughout the
paper we set $c=1$. The exterior derivative of the 1-form $A$ is the Faraday 2-form $F$ \cite{ryder} where
\begin{eqnarray}
\label{faraday}
F = dA &=& -\frac{1}{2} F_{\mu \nu} dx^\mu \wedge dx^\nu  \nonumber \\
&=& (E_x dx + E_y dy + E_z dz ) \wedge dt + B_x dy \wedge dz + B_y dz \wedge dx + B_z dx \wedge dy  ~.
\end{eqnarray}
We have written the Faraday 2-form in 4-vector and 3-vector notation. In \eqref{faraday} we have used
the wedge products like $dx \wedge dt$, $dy \wedge dz$ which are anti-symmetric under exchange of the differentials.

With this differential form notation one can write down the expression for the usual static, magnetic
AB phase shift. In the case where one has a solenoid with a static current and magnetic field, the 4-vector
potential becomes $A_\mu = (0, {\bf A})$, {\it i.e.} the scalar potential is zero. Thus the phase AB shift, 
$\delta \alpha _{{\rm AB}}$, for this case \cite{AB, ES} becomes
\begin{equation}
\label{phase1}
\delta \alpha _{{\rm AB}} = \frac{e}{\hbar} \oint {\bf A} \cdot d{\bf x} = \frac{e}{\hbar} \int {\bf B} \cdot d{\bf S} ~, 
\end{equation}  
where $e$ is the charge of the particle. In arriving at \eqref{phase1} we have used Stokes' theorem
\eqref{stokes-df} and the fact that $B_x dy \wedge dz + B_y dz \wedge dx + B_z dx \wedge dy = {\bf B} \cdot d {\bf S}$.
Note that the AB phase shift, $\delta \alpha _{{\rm AB}}$, can be written equivalently as either the closed path
integral of ${\bf A}$ or the surface integral of the magnetic field, {\it i.e.} the curl of ${\bf A}$. 
\section{Time-dependent Aharonov-Bohm effect}
Now we use the results of the previous section to address the case when the current and magnetic flux through 
the solenoid vary with time. To this end we need to write down the 4-vector potential for the solenoid.
The 3-vector potential inside and outside the solenoid (which is taken to have a radius $\rho = R$) is
\begin{eqnarray}
\label{3-vector-a}
{\bf A}_{{\rm in}} &=& \frac{\rho B(t)}{2} \hat{\bf \varphi} ~~~~{\rm for ~~ \rho <R} \nonumber \\
{\bf A}_{{\rm out}} &=& \frac{B(t) R^2}{2 \rho} \hat{\bf \varphi} ~~~~{\rm for ~~ \rho \ge R} ~,
\end{eqnarray}
and the scalar potential is normally taken as zero everywhere {\it i.e.} $\phi = 0$. We will come back to this gauge 
choice of $\phi =0$ shortly. The magnetic field, $B(t)$, now depends on time since the current of the solenoid
is being varied. We could write this time dependent magnetic field in terms of the time dependent current
through the solenoid, $I(t)$, and the number of turns per unit length of the solenoid, $N$, as 
$B(t) \propto N I(t)$. However for our purposes the time dependence can just be left in terms of the 
time variation of the amplitude of the magnetic field. Taking the curl of the 3-vector potential 
in \eqref{3-vector-a} yields the magnetic field
\begin{eqnarray}
\label{B-field}
{\bf B}_{{\rm in}} &=& \nabla \times {\bf A}_{{\rm in}} =  B(t) {\bf {\hat z}}~~~~{\rm for ~~ \rho <R} \nonumber \\
{\bf B}_{{\rm out}} &=& \nabla \times {\bf A}_{{\rm out}} = 0  ~~~~{\rm for ~~ \rho \ge R} ~,
\end{eqnarray}
the only difference from the static case is now the magnetic field inside the solenoid is time varying.
The magnetic field outside is still zero. The new feature resulting from allowing the magnetic flux to
vary with time is that there is an electric field coming from ${\bf E} = - \partial _t {\bf A}$. 
Explicitly using \eqref{3-vector-a} we find
\begin{eqnarray}
\label{E-field}
{\bf E}_{{\rm in}} &=& - \frac{\partial {\bf A}_{{\rm in}}}{\partial t} = - \frac{\rho {\dot B(t)}}{2} \hat{\bf \varphi}  
~~~~{\rm for ~~ \rho <R} \nonumber \\
{\bf E}_{{\rm out}} &=& - \frac{\partial {\bf A}_{{\rm out}}}{\partial t} =  
- \frac{{\dot B(t)} R^2}{2 \rho} \hat{\bf \varphi}
~~~~{\rm for ~~ \rho \ge R} ~,
\end{eqnarray}
where the overdots are derivatives with respect to time. One point to note is that while \eqref{3-vector-a}
\eqref{B-field} \eqref{E-field} assume arbitrary time dependence for the flux, $B(t)$, there is actually some
restriction coming from Maxwell's equations that must be taken into account. While Faraday's Law 
$\nabla \times {\bf E} = - \partial _t {\bf B}$ is consistent with an arbitrary time dependence for
$B(t)$, the sourceless Ampere-Maxwell equation, $\nabla \times {\bf B} = \partial _t {\bf E}$ only
works with the expressions in equations \eqref{B-field} and \eqref{E-field} if the flux is linearly 
dependent on time {\it i.e.}
\begin{equation}
\label{bt}
B(t) = B_0 + B_1 t ~,
\end{equation}
where $B_0 , B_1$ are constants. Later in section \eqref{sin-flux} we will show that it is possible to
consider other variations of flux other than the linear dependence of \eqref{bt}. This is accomplished by changing the
spatial dependence from $\rho$ ($\frac{1}{\rho}$) for the fields inside (outside) the solenoid. For the
sinusoidal dependence considered in section \eqref{sin-flux} we find that the spatial dependence will be 
ordinary Bessel functions. However for this section it is good to keep in mind that strictly the time
dependence of the flux is that given in equation \eqref{bt}. It turns out that this linear time dependence, although
unphysical for arbitrary times (there is some practical limit to how large of a B-field one can make)
it is good for illuminating the unique features of the time dependent Aharonov-Bohm effect.

We now evaluate the AB phase shift using the surface area integral of the fields {\it i.e.} $\int _c F $
where $F$ is the Faraday 2-form \eqref{faraday}. The difference from the static case, aside from
the time variation of the magnetic field in \eqref{B-field}, is that there is a time dependent 
electric field \eqref{E-field} which contributes to the phase. The AB phase in terms of the Faraday
2-form ({\it i.e.} in terms of the electric and magnetic fields) is
\begin{eqnarray}
\label{phase2}
\delta \alpha _{{\rm AB}} &=& \frac{e}{\hbar} \int _c F = \frac{e}{\hbar}  \int (E_x dx + E_y dy + E_z dz ) \wedge dt 
+ B_x dy \wedge dz + B_y dz \wedge dx + B_z dx \wedge dy  \nonumber \\
&=& - \frac{e}{\hbar} \oint {\bf A} \cdot d{\bf x} + \frac{e}{\hbar}\int {\bf B } \cdot d {\bf S} = 
- \frac{e}{\hbar} \oint {\bf A} \cdot d{\bf x} + \frac{e}{\hbar} \oint {\bf A} \cdot d{\bf x} = 0~.
\end{eqnarray} 
In the second line we have written the three electric field terms from the first line as  
$- \oint {\bf A} \cdot d{\bf x}$ using ${\bf E} = - \partial _t {\bf A}$ and performing the
$dt$ integration. Next the three magnetic terms of the first line were converted from differential
form notation to 3-vector notation, $\int {\bf B } \cdot d {\bf S}$. Then using 
${\bf B} = \nabla \times {\bf A}$ and the 3-vector form of Stokes' theorem we
end up with $+ \oint {\bf A} \cdot d{\bf x}$. Thus we find that the electric and
magnetic contributions in \eqref{phase2} exactly cancel giving the prediction of
{\it no time-dependent phase shift} for the time-dependent Aharonov-Bohm effect. We should clarify 
that this cancellation is only in effect for the time-dependent part of the magnetic field.
If the vector potential can be split into time-independent and time-dependent parts 
${\bf A} = {\bf A}_0 ({\bf x}) + {\bf A}_1 ({\bf x , t})$ then the magnetic field also splits into
time-independent and time-dependent parts ${\bf B}= \nabla \times {\bf A} = {\bf B}_0 ({\bf x}) + {\bf B}_1 ({\bf x , t})$.
The electric field only comes from  ${\bf A}_1 ({\bf x , t})$ and thus only the time-dependent magnetic
field ${\bf B}_1 ({\bf x , t})$ part is canceled by the electric field. Initially one might expect that the AB phase
in \eqref{phase2} would be time-dependent with the time-dependence coming from the time-varying
magnetic flux. Instead, we find that, at least according to the Faraday 2-form/E\& M field
expression for the AB phase, that the AB phase is time-independent due to the cancellation
of the electric and magnetic field parts. This is exactly what was found in  \cite{singleton} where
it was shown that the standard time-dependent AB phase shift due to the magnetic field was canceled
by the phase shift coming from the electric force on the electrons. The electric field would
accelerate/decelerate the electrons thus changing their position with respect to the case when there
was no electric field outside the solenoid {\it i.e.} for a static magnetic flux. This shift in position, 
associated with the electric field, would give a corresponding shift in phase.  

Since we are considering time dependence of the flux one encounters the questions of ``what surface?" 
and ``at what time?" are meant in equation \eqref{phase2}. To address these questions we
look at the change in phase along some infinitesimal path length $\Delta {\bf x}$ and some associated infinitesimal
area  $\Delta {\bf S}$, and show that for these infinitesimal temporal and geometrical quantities that one gets
cancellation between the electric contribution in \eqref{phase2} and the {\it time dependent} contribution of the
magnetic field. Thus adding/integrating these infinitesimal quantities up leads to the result in equation 
\eqref{phase2}. For an infinitesimal time interval, $\Delta t$, length interval $\Delta {\bf }$ and area interval, 
$\Delta {\bf S}$ one can write \eqref{phase2} as
\begin{eqnarray}
\label{phase2a}
\Delta (\delta \alpha_{{\rm AB}} ) &=& \frac{e}{\hbar} \left( {\bf E} \cdot \Delta {\bf x} \Delta t +
\Delta {\bf B} \cdot \Delta {\bf S} \right) \nonumber \\ 
&=& \frac{e}{\hbar} \left( - \frac{R^2}{2 \rho} (\dot B \Delta t)(\rho \Delta \varphi ) 
+ \frac{\Delta \varphi R^2}{2} (\dot B \Delta t) \right) = 0 ~.
\end{eqnarray}
In \eqref{phase2a} we have Taylor expanded $B(t) = B_0 + {\dot B} \Delta t + {\cal O} (\Delta t )^2$. The
$B_0$ term gives the usual, static part of the Aharonov-Bohm phase shift; the focus of
this paper is the time-dependent contribution which to first order in $\Delta t$ is given by
the second term in the Taylor expansion for $B(t)$. In \eqref{phase2a} we have also
used $\Delta {\bf x} = (\rho \Delta \varphi ) {\hat \varphi}$, 
$\Delta {\bf S} =\frac{1}{2}(R \Delta \varphi )R {\hat z}= \frac{1}{2}\Delta \varphi R^2 {\hat z}$, and
the expression for ${\bf E}_{{\rm out}}$ from \eqref{E-field}. Adding up/integrating
all these infinitesimal phase shifts, with $\Delta (\delta \alpha_{{\rm AB}} ) = 0$, gives the result
$\delta \alpha_{{\rm AB}}=0$ in equation \eqref{phase2} {\it i.e.} the time-dependent part of the phase shift cancels.
The cancellation between the electric and magnetic contributions to the phase in
\eqref{phase2a} is made more transparent in the case when the flux varies linearly as in \eqref{bt}.
In this case one can replace $\Delta t$ by a finite time interval $T$ and the electric contribution in 
\eqref{phase2a} takes the form
$$
- \frac{e}{\hbar} \left( \frac{R^2}{2 \rho} (B _1 T)(\rho \Delta \varphi ) \right)= 
- \frac{e}{\hbar} \left( \frac{R^2 \Delta \varphi }{2} (B _1 T) \right)~.
$$
The magnetic contribution takes the form
$$
+ \frac{e}{\hbar}\left( \frac{R^2 \Delta \varphi }{2} (B_0 + B_1 T) \right) ~.
$$
Comparing these two expressions one finds that the time dependent pieces ({\it i.e.} 
the terms $ \propto B_1 T$) cancel leaving only the static contribution ({\it i.e.} the term $\propto B_0$).

However the above analysis, as well as that in \cite{singleton}, leaves open the question as to 
the status of Stokes' theorem in the time-dependent case in regard to the one-form side of
equation \eqref{stokes-df}. At first glance it would seem Stokes' theorem is
violated in this case which would then call into question the above analysis. In the above we have calculated
the AB phase $\delta \alpha _{{\rm AB}}$, through the right hand side of Stokes' theorem as given in \eqref{stokes-df} 
using the Faraday 2-form $d\omega = dA  = F$ from  \eqref{faraday}. The result was $\delta \alpha _{{\rm AB}} =0$.
We now calculate the left hand side of Stokes' theorem as given in \eqref{stokes-df}
using the 4-vector potential 1-form $\omega = A$ \eqref{a-form} and at first find that, apparently, 
$\delta \alpha _{{\rm AB}} \ne 0$ which would imply a violation of the 4-dimensional Stokes' theorem. In the end
we resolve this through the non-simply connected topology of the Aharonov-Bohm set-up and the 
associated non-single valued gauge potentials.

The 4-vector potential in this case is given by
$A^\mu = (\phi , {\bf A})  = (0,  {\bf A} (t, {\bf x}))$ where the 3-vector potential is given by 
\eqref{3-vector-a} and the scalar potential is zero. This {\it gauge} choice for $A^\mu$  does give 
the correct magnetic \eqref{B-field} and electric fields \eqref{E-field} for this situation. But we will
see that there are other gauges for which the value of $\int _c \omega = \int A$ will depend on the gauge
due to the non-single valued nature of the gauge transformation function which is connected with the 
fact that the space in this case is {\it non-simply connected} (see page 102-103 of \cite{ryder} for a discussion
on this point). Using $\phi =0$ and ${\bf A} (t, {\bf x})$ from \eqref{3-vector-a} we apparently find that the 
AB phase shift in terms of $A^{\mu}$ is
\begin{eqnarray}
\label{phase3}
\delta \alpha _{{\rm AB}} &=& -\frac{e}{\hbar} \int _{\partial c} \omega = -\frac{e}{\hbar} \oint A_\mu dx^\mu = 
 -\frac{e}{\hbar} \left( \int \phi dt - \oint {\bf A} (t, {\bf x}) \cdot d {\bf x} \right) 
=  \frac{e}{\hbar} \oint {\bf A} (t, {\bf x}) \cdot d {\bf x} \nonumber \\
&=& \frac{e}{\hbar} \int \nabla \times{\bf A} (t, {\bf x}) \cdot d {\bf S} = 
\frac{e}{\hbar} \int {\bf B} (t, {\bf x}) \cdot d {\bf S} ~,
\end{eqnarray}
where in going from the end of the first line to the second line we have used the Stokes' theorem in
the form \eqref{stokes3d} and then used $\nabla \times {\bf A} = {\bf B}$. The time dependence of the AB phase
shift in \eqref{phase3} is exactly same as the magnetic part of the AB phase shift given in
\eqref{phase2} where the phase shift was calculated using the Faraday 2-form. Thus from the 4-vector
potential calculation in \eqref{phase3} it appears that we should 
get the usual magnetic AB phase shift but with a time dependence coming from the time dependence of the 
magnetic field. This is what was predicted in earlier work on the time-dependent AB effect \cite{chiao}.
But in this way one does not take into account the effect of the electric field. In the work \cite{stern}
time dependent Berry Phases were briefly considered and it was noted that in the case of time-dependent 
geometric fluxes that there would be a ``motive force" similar to the electromotive force in Faraday's law.
However in this paper we emphasize that our analysis shows an exact cancellation of the ``magnetic" and ``electric" 
contributions to the time-dependent part of the AB phase shift. 

To get an ``electric" and ``magnetic" cancellation of the potentials, $\phi$ , ${\bf A}$ one would need a non-zero
scalar potential. In fact one {\it can} find a gauge transformation which does give scalar and 3-vector potentials
which give the magnetic and electric fields from \eqref{B-field} \eqref{E-field} but for which $\phi \ne 0$. 
We begin by noting that one can get the outside magnetic and electric fields from \eqref{B-field} \eqref{E-field}
by taking the scalar potential $\phi _{{\rm out}} = R^2 {\dot B}(t) \varphi /2$ and ${\bf A}_{{\rm out}} = 0$. Taking
${\bf E}_{{\rm out}} = -\nabla \phi _{{\rm out}} - \partial_t {\bf A}_{{\rm out}}$ does give 
${\bf E}_{{\rm out}} = - \frac{{\dot B(t)} R^2}{2 \rho} \hat{\bf \varphi}$  
and taking ${\bf B}_{{\rm out}} = \nabla \times {\bf A}_{{\rm out}}$ does give ${\bf B}_{{\rm out}} =0$.
It will be noticed that the scalar potential is non-single valued due to the presence of the angular
coordinate $\varphi$-dependence. Such non-single valued functions can not exist (or are pathological) in simply 
connected spaces, but the Aharonov-Bohm setup is {\it non-simply} connected. Because of this
non-single value functions can be considered (see the discussion on page 102-103 of \cite{ryder}). The
non-single valued form of the potentials is connected to the form of the gauge potentials 
given in \eqref{3-vector-a} by the following gauge transformation 
\begin{equation}
\label{chi}
A^\mu \rightarrow A^\mu + \partial ^\mu \chi ~~~;~~~ \chi =\frac{K R^2 B(t) \varphi}{2} ~,
\end{equation}
where $K$ is some constant in the range $0 \le K \le 1$. Note that the gauge function $\chi$
is also non-single valued and this is again connected with the fact that the space in the 
Aharonov-Bohm setup has non-simply connected topology. With this gauge transformation \eqref{chi}
the outside gauge potentials from \eqref{3-vector-a} become 
\begin{equation}
\label{4-vector-a}
\phi _{{\rm out}} = K\frac{R^2 {\dot B}(t) \varphi}{2} ~~~;~~~ {\bf A}_{{\rm out}} = (1-K)\frac{R^2 B(t)}{2 \rho} {\hat {\bf \varphi}} ~.
\end{equation} 
When $K=0$ the outside electric field is given purely in terms of the non-zero, single-valued 3-vector potential, and
when $K=1$ the outside electric field is given purely in terms of the non-zero, non-single-valued scalar potential. For $K$
at intermediate values the outside electric field ${\bf E} = -\nabla \phi _{{\rm out}} - \partial _t {\bf A}_{{\rm out}}$
comes from a combination of the scalar and 3-vector potential. In all cases the outside magnetic field is zero.
One important thing to point out is that the quantity $\oint A_\mu dx^\mu$ is no longer gauge invariant
due to the non-single valued character of the gauge function $\chi$ from \eqref{chi} which is related to the
non-simply connected topology of the Aharonov-Bohm set-up. Explicitly under the gauge transformation \eqref{chi}
we find
\begin{equation}
\label{gauge-trans}
\oint A_\mu dx^\mu \rightarrow \oint A_\mu dx^\mu + \oint \partial _\mu \chi dx^\mu = \oint A_\mu dx^\mu + (\chi (f) - \chi (i)) ~,
\end{equation}  
where $\chi (f) - \chi (i)$ is the difference between the final and initial point of the path. For a closed path and
a single value $\chi$ this will be zero and $\oint A_\mu dx^\mu$ will be gauge invariant. However for a non-single
value gauge function $\chi (f) - \chi (i)$ will not be zero for a closed path and 
$\oint A_\mu dx^\mu$ is {\it not} gauge invariant. Note that the AB phase shift as given in terms of the
Faraday 2-form \eqref{phase2} is still gauge invariant since the electric and magnetic fields 
as well as the ``area" $dx^\mu \wedge dx^\nu$ are the same regardless of whether the fields
come from a single valued 3-vector potential as in \eqref{3-vector-a}, a non-single valued scalar potential
$\phi_{{\rm out}} =R^2 B(t) \varphi /2$ or some combination of the two as in \eqref{4-vector-a}. The question arises is there
a gauge ({\it i.e.} a choice of $K$ in \eqref{chi}) for which 
$\oint A_\mu dx^\mu =  -\frac{1}{2} \int F_{\mu \nu} dx^\mu \wedge dx^\nu$ ?
The answer is ``yes" for $K=\frac{1}{2}$. For this choice we have
\begin{equation}
\label{4-vector-a2}
\phi _{{\rm out}} = \frac{R^2 {\dot B}(t) \varphi}{4} ~~~;~~~ {\bf A}_{{\rm out}} = \frac{R^2 B(t)}{4 \rho} {\hat {\bf \varphi}} ~.
\end{equation}
For this gauge choice of the potentials the electric field is seen to come equally from $\phi _{{\rm out}}$ 
and ${\bf A}_{{\rm out}}$
\eqref{4-vector-a2} 
$$
{\bf E} = -\nabla \phi _{{\rm out}} - \partial _t {\bf A} _{{\rm out}} = 
-\frac{R^2 {\dot B}(t)}{4 \rho} - \frac{R^2 {\dot B}(t)}{4 \rho} = - \frac{R^2 {\dot B}(t)}{2\rho} ~.
$$
For the gauge choice, $K=1/2$, one can see that the contributions of $\phi _{{\rm out}}$ and ${\bf A}_{{\rm out}}$ to
$\oint A_\mu dx^\mu$ cancel thus bringing the gauge potential expression for the AB phase shift into agreement
with the Faraday 2-form expression for the AB gauge. For the interval $\Delta t$ the infinitesimal
contributions from the scalar and vector potentials in \eqref{4-vector-a2} become
\begin{eqnarray}
\label{int-gauge}
\Delta \phi \Delta t - \Delta {\bf A} \cdot \Delta {\bf x} \nonumber = 
\frac{R^2 {\Delta B} \Delta \varphi}{4 \Delta t} \Delta t - 
\frac{R^2 \Delta B}{4 \rho} \rho \Delta \varphi = 0 ~, 
\end{eqnarray} 
where we have used  $d{\bf x} \rightarrow (\rho \Delta \varphi ) {\hat \varphi}$. Adding up ({\it i.e.} integrating)
these infinitesimal contributions from \eqref{int-gauge} gives $\oint A_\mu dx^\mu = 0$ and we
find that the 4-dimensional Stokes' theorem ({\it i.e.} 
$\oint A_\mu dx^\mu = -\frac{1}{2} \int F_{\mu \nu} dx^\mu \wedge dx^\nu$)
is now satisfied but only for the gauge $K=\frac{1}{2}$. In the arguments above, this gauge dependence of the 
4-dimensional Stokes' theorem can be traced to the gauge dependence of $\oint {\bf A} _\mu dx ^\mu$ which
arises from the non-single valued character of the gauge transformation function $\chi$ in \eqref{chi}. This in turn is
connected with the non-simply connected topology of the Aharonov-Bohm set-up. We note that the Faraday 2-form
side of the 4-dimensional Stokes' theorem, $ -\frac{1}{2} \int F_{\mu \nu} dx^\mu \wedge dx^\nu$, is gauge
invariant even for the non-single valued gauge function $\chi$ from \eqref{chi}. Thus, one can experimentally 
test the correctness (or not) of the above arguments. If one performs the time-dependent AB experiment and finds
that the phase shift, $\delta \alpha _{{\rm AB}}$, is not time-dependent then the above arguments are correct; 
if one performs the time-dependent AB experiment and finds that the phase shift, $\delta \alpha _{{\rm AB}}$, inherits
the time dependence of the magnetic field/magnetic flux then the above arguments are not correct. However if this
last case is the one selected by experiment one needs to understand why the electric field in the time-dependent case
has no influence on the phase shift as given by \eqref{phase2}.   
\section{Sinusoidal flux variation}
\label{sin-flux}
Although up to now we have assumed an arbitrary time variation for the flux, $B(t)$, as already mentioned,
due to the restriction coming from Maxwell-Ampere's Law ({\it i.e.} $\nabla \times {\bf B} = \partial _t {\bf E}$)
one is implicitly dealing with only a linearly increasing flux as given in equation \eqref{bt}. While
this linearly increasing flux is good for illustrating the basic features of the time dependent Aharonov-Bohm
effect (in particular the claimed cancellation between the electric and magnetic contributions to the
phase shift given in \eqref{phase2} or \eqref{phase2a}) one might ask if more general time variations 
can be considered, and if so does one still have the same cancellations of the electric and magnetic contributions to the
phase shift. In this section we show this is possible for the physically realistic case of sinusoidally varying fields.
The reason to focus on sinusoidally varying fields is that they would be the ones most likely used experimentally
to test the predictions made in this paper. To begin we will assume a more general $\rho$ dependence for
the vector potential. Previously in \eqref{3-vector-a} we had taken the $\rho$ dependence of ${\bf A}$ as $\propto \rho$
and $\propto 1/ \rho$ for inside and outside the solenoid respectively. Here we assume a vector potential of the form
\begin{equation}
\label{3-vector-at}
{\bf A} = F (\rho) e^{i \omega t} \hat{\bf \varphi} ~,
\end{equation}
where $F(\rho)$ is some function of $\rho$ and we have already put in the assumed sinusoidal time dependence
with frequency $\omega$. The direction of ${\bf A}$ is still taken to be in the $\hat{\bf \varphi}$ direction. 
Using this form of the vector potential in \eqref{3-vector-at} to calculate the magnetic and electric fields via
${\bf B} = \nabla \times {\bf A}$, ${\bf E} = - \partial _t {\bf A}$ and then inserting
these in $\nabla \times {\bf B} = \partial _t {\bf E}$ we arrive at the following equation for $F(\rho)$
\begin{equation}
\label{Fr}
F '' (\rho) + \frac{F' (\rho)}{\rho} - \frac{F(\rho)}{\rho ^2} + \omega ^2 F (\rho) = 0 ~,
\end{equation}
where primes denote differentiation with respect to $\rho$. Changing variables to $x = \omega \rho$ the
equation \eqref{Fr} becomes 
\begin{equation}
\label{Fx}
F '' (x) + \frac{F' (x)}{x} + \left( 1 - \frac{1}{x ^2} \right) F (\rho) = 0 ~,
\end{equation}
where now the primes denote differentiation with respect to $x$. Equation \eqref{Fx} is solved by
the ordinary Bessel functions of order $1$ namely $J_1 (x)$ and $Y_1 (x)$. The solutions for the
vector potentials inside and outside the solenoid now take the form 
\begin{eqnarray}
\label{3-vector-a2}
{\bf A}_{{\rm in}} &=& A_1 J_1 (\omega \rho ) e^{i \omega t} \hat{\bf \varphi} ~~~~{\rm for ~~ \rho <R} \nonumber \\
{\bf A}_{{\rm out}} &=& \left[ C_1 J_1 (\omega \rho ) + D_1 Y_1 (\omega \rho ) \right] 
e^{i \omega t} \hat{\bf \varphi} ~~~~{\rm for ~~ \rho \ge R} ~,
\end{eqnarray}
where $A_1 , C_1 , D_1$ are constants to be determined from boundary conditions. Note that ${\bf A}_{{\rm in}}$
only uses $J_1 (\omega \rho )$ since $Y_1 (\omega \rho )$ diverges at $\rho =0$. From the vector potential 
in \eqref{3-vector-a2} one can calculate the ${\bf E}$ and ${\bf B}$ fields. 
Using these new ${\bf E}$ and ${\bf B}$ fields and repeating the arguments leading to the results in \eqref{phase2} 
or \eqref{phase2a}, it is straight forward to verify that the electric and magnetic contributions 
to the {\it time dependent} parts of the Aharonov-Bohm phase shift still cancel. 
Different time dependences for the flux will lead to different forms for $F(\rho )$
but the linear dependence considered previously in \eqref{bt} and the sinusoidal dependence considered in this
section are the most interesting cases from a theoretical point of view and from an experimental point of view.
The linear case is interesting since it most clearly illustrates the cancellation of the time dependent part of the
Aharonov-Bohm phase shift with very little approximation. Also for a fixed time interval one can arrange 
to have a linearly increasing flux. The sinusoidal case is probably the easiest situation to set up experimentally
and it would most dramatically demonstrate the effect predicted in this work -- if the above analysis is correct
the Aharonov-Bohm interference pattern should not (contrary to earlier expectations) shift in time to the frequency of the
sinusoidally varying flux. 

One final point to make in this case  of a sinusoidally varying flux is that now there is a non-zero magnetic field
outside the solenoid. In the previous case using ${\bf A}_{{\rm out}}$ from \eqref{3-vector-a} one got 
${\bf B}_{{\rm out}} = \nabla \times {\bf A}_{{\rm out}} = 0$. However using ${\bf A}_{{\rm out}}$ from \eqref{3-vector-a2}
gives ${\bf B}_{{\rm out}} = \nabla \times {\bf A}_{{\rm out}} \ne 0$. Thus for a sinusoidally varying field there are both
${\bf E}$ and ${\bf B}$ fields outside the solenoid which makes this conceptually different from the usual
time independent Aharonov-Bohm effect. However as mentioned above the time dependent parts of the electric and
magnetic contributions from \eqref{phase2} or \eqref{phase2a} still cancel. 
\section{Multivalued monopole potential}
There is another situation where one encounters non-single valued potentials and gauge transformation functions --
the potentials for a magnetic monopole. The following 3-vector potential (now using spherical polar
coordinates $r, \theta, \varphi$, rather than the cylindrical coordinates, $\rho, \varphi, z$ used in
the previous section) 
\begin{equation}
\label{monopole}
{\bf A} _{{\rm monopole}} = \frac{g (1-\cos \theta )}{r\sin \theta} {\hat {\bf \varphi}} ~,
\end{equation} 
yields a monopole magnetic field ${\bf B} = \nabla \times {\bf A} = g {\hat {\bf r}}/ r^2$. The vector potential
in \eqref{monopole} is single valued, but it has the usual Dirac string singularity pathology along the negative
z-axis {\it i.e.} $\theta = \pi$. One can also obtain a magnetic monopole field
from ${\bf A} _{monopole} = - \frac{g (1+\cos \theta )}{r\sin \theta} {\hat {\bf \varphi}}$ which has a
Dirac string singularity along the positive z-axis {\it i.e.} $\theta = 0$. These two forms of the monopole
3-vector potential are related by the gauge transformation ${\bf A} \rightarrow {\bf A} - \nabla \chi$
with $\chi = 2 g \varphi$. In this case the gauge function $\chi$ is non-single valued but 
the two forms of the gauge potential ${\bf A} _{{\rm monopole}}$ are single valued. Thus, this is 
not exactly like the time-dependent AB effect of the previous section. 

It is easy to see that one can also get a magnetic monopole  field from the following, alternative 3-vector 
potential \cite{arfken} \cite{pal}
\begin{equation}
\label{monopole1}
{\bf A} _{{\rm monopole}} = - \frac{g \varphi \sin \theta }{r} {\hat {\bf \theta}} ~,
\end{equation} 
which does not have Dirac string singularity of \eqref{monopole} but it is non-single valued 
due to the $\varphi$ dependence of $A_\theta$. The two vector potentials in 
\eqref{monopole} and \eqref{monopole1} are related by a gauge transformation of the form
\begin{equation}
\label{mono-gauge}
A^\mu \rightarrow A^\mu + \partial ^\mu \chi ~~~;~~~ \chi = - g (1 - \cos \theta) \varphi
\end{equation}
Here we see that both the gauge transformation function $\chi$ in \eqref{mono-gauge} {\it and} the
3-vector gauge potential \eqref{monopole1} are non-single valued, which then is similar to the time-dependent AB
effect of the previous section. In this work we just point out the similarity between the time-dependent
AB effect and magnetic monopoles. We will return to a detailed analysis of the monopole case in 
future work.   
\section{Summary}
In this paper we have investigated the time-dependent AB effect and related issues connected with 
Stokes' theorem and gauge symmetry. We found that the 4-dimensional Strokes' theorem as given by 
\eqref{stokes-df} with $\omega =A$ and $d \omega = dA =F$ is gauge {\it dependent} since 
$\oint _{\partial c} A$ is gauge dependent. This comes about due to the non-single valued character of
scalar potential $\phi$ from \eqref{4-vector-a} and the non-single valued character of the gauge function
$\chi$ from \eqref{chi}. This non-single valued character of $\phi$ and $\chi$ are the result of the 
topology of the Aharonov-Bohm set-up being non-simply connected \cite{ryder}. Since the quantity 
$\oint _{\partial c} A$ is gauge dependent the equality of the left and right hand sides of 
Stokes' theorem given in \eqref{stokes-df} will only be true for a specific gauge. In terms of
the scalar and 3-vector potentials given in \eqref{4-vector-a} the gauge where the left hand
side and right hand side of Stokes' theorem are equal is $K=1/2$. This analysis can be tested experimentally
by performing the time-dependent Aharonov-Bohm experiment. If one finds that $\delta \alpha _{{\rm AB}}$ does
not inherit the time-variation of the magnetic flux then the above analysis is correct; if $\delta \alpha _{{\rm AB}}$
inherits the time dependence of the magnetic field then the above analysis is not correct. 

We concluded by showing that some of these issues of non-single valuedness of the gauge potentials
and gauge transformation function also appear in the analysis of magnetic monopoles. We will return to
a more detailed examination of this in future work.
{\par\noindent {\bf Acknowledgments:}}
DS thanks Sodsri Singleton for her support during the period when this work was completed. JM acknowledges 
the support of Donna Ramacciotti.   

\end{document}